\documentclass[12pt,letterpaper]{article}
\pdfoutput=1
\usepackage[includeheadfoot,
            marginratio={1:1,2:3}, 
            width=412pt, 
            height=688pt,]{geometry}

\usepackage{amsmath}
\usepackage{amsfonts}
\usepackage{amssymb}
\usepackage{graphicx}
\usepackage{cite}
\usepackage{ulem}
\usepackage{dsfont}
\usepackage{longtable}
\usepackage{afterpage}
\usepackage{booktabs}
\usepackage{hyperref}
\usepackage{tikz} 
\usetikzlibrary{decorations.pathreplacing}
\usepackage{subfigure}
\usepackage{verbatim}
\usepackage{multirow}
\usepackage{physics}
\usepackage{braket}
\usepackage{breqn}

\newcommand{\ov}{\overline}

\allowdisplaybreaks[2]
\numberwithin{equation}{section}

\begin{document}

\normalem
\vspace*{-1.5cm}
\begin{flushright}
  {\small
  MPP-2019-97\\

  }
\end{flushright}

\vspace{1.5cm}

\begin{center}
  {\LARGE
Closed Bosonic String Tachyon Potential from the $\mathcal{N}=1$ Point of View \\[0.3cm]    
}
\vspace{0.4cm}

\end{center}

\vspace{0.35cm}
\begin{center}
 Lorenz Schlechter
\end{center}

\vspace{0.1cm}
\begin{center} 
\emph{Max-Planck-Institut f\"ur Physik (Werner-Heisenberg-Institut), \\ 
   F\"ohringer Ring 6,  80805 M\"unchen, Germany } \\[0.1cm] 
\vspace{0.25cm}

\vspace{0.2cm}

\end{center} 

\vspace{1cm}


\begin{abstract}
\noindent
The potential of the closed bosonic bulk tachyon is investigated using $\mathcal{N}~=~1$ closed superstring field theory formulated around Berkovits' and Vafas embedding of the bosonic string into the superstring. From this point of view the bosonic minimum gets destabilized by new deformations of the underlying CFT. Another critical point of the potential is identified by solving the equation of motions and the properties of this new solution are discussed.
\end{abstract}


\clearpage
\tableofcontents


\section{Introduction}
A long standing mystery of string theory is the closed bosonic string bulk tachyon. Unlike the open string tachyon, which is well understood to describe the decay of D-branes, the process behind the closed string tachyon is much less understood. Localized closed string tachyons are known to describe the decay of the geometric backgrounds they are localized on. For example the decay of $\mathbb{C}/\mathbb{Z}_n$ orbifolds to flat spacetimes \cite{Adams:2001sv}, the decay of twisted circles\cite{David:2001vm} or the decay of compact spacetimes into nothing\cite{Headrick:2004hz}. Using RG-flow techniques, it was shown that tachyon condensations in super- and subcritical string theories lead back to the critical 10d string theories \cite{Suyama:2005wd}. Also some exact solutions interpolating between different heterotic and between Type 0 and Type II superstrings were found \cite{Hellerman:2004zm,Hellerman:2004qa,Hellerman:2006ff, Hellerman:2006hf}. These solutions are based on a non-zero dilaton gradient and assume an either light-like or time-like tachyon profile.

An analogous understanding of the bosonic bulk tachyon is still lacking. One conjecture is that it leads to the supersymmetric 10d vacua and therefore describes the decay of 16 dimensions \cite{West:2002hh}. This would require the spontaneous breaking of the 26-dimensional Lorentz invariance to a 10-dimensional Lorentz invariance. In \cite{Lust:1987ik} it was shown that the bosonic string compactified on 16-dimensional torus with appropriate boundary conditions is equivalent to heterotic or superstring theories. 

The potential of the closed bosonic bulk tachyon was first computed in \cite{PhysRevD.42.1289} up to the massless level and qubic order
\begin{equation}
V=-t^2+{6561\over 4096}t^3\;,
\end{equation}
where t is the tachyon field. There is a non-perturbative minimum at $t\approx 0.416$. But closed string field theory is non-polynomial. The addition of the quartic tachyon contact-term, $-3.0172t^4$, seemed to destroy this minimum\cite{Moeller:2004yy,Belopolsky:1994bj}. The inclusion of the massless fields, especially the dilaton, leads to a flattening of the potential. It was conjectured that the endpoint of tachyon condensation could be a zero action minimum describing a state without any dimensions\cite{Yang:2005rx}. But a level 10 computation up to quartic order showed that the minimum seems to converge to a fixed value of $-0.05$ \cite{Moeller:2006cv}. Also the quintic tachyon term was computed to be $(9.924\pm 0.008)t^5$\cite{Moeller:2006cw}.

This paper aims to analyze the same problem, the minimum of the bosonic bulk tachyon potential in string field theory, in a different setting. Instead of the bosonic string field theory, $\mathcal{N}=1$ superstring field theory is applied, using Berkovit's embedding of the bosonic string in the superstring moduli space \cite{Berkovits:1993xq}.  If the conjecture that the bulk tachyon leads to the critical superstring theories is correct, the result should be a new minimum representing the critical superstring vacuum. In this case the bosonic string field theory vacuum is a false vacuum, lying at the boundary of the bosonic moduli space. It should be destabilized by new string field components which are absent in the purely bosonic case and therefore only be a saddle point of the whole potential.

This paper is organized as follows. In section $2$ the world sheet construction of Berkovits and Vafa \cite{Berkovits:1993xq} as well as the construction of closed superstring field theory is reviewed. In section 3  the potential is evaluated. We identify the map between the bosonic states and the states in the $\mathcal{N}=1$ theory and show how the bosonic potential arises explicitly. The minima are no longer critical points of the potential but get destabilized by new deformations. In section $4$ we find some solutions to the equations of motions (EOM), identifying new critical points. Finally in section 5 the interpretation of these solutions as well as possible further research directions are discussed.
\section{Preliminaries}
String field theory is formulated around a background CFT. The string field $\Psi$ consists of all possible deformations of the theory, i.e. all states of the theory. We start by describing a very special CFT which was shown by Berkovits to give an embedding of the bosonic string in the superstring \cite{Berkovits:1993xq}.

\subsection{The World-sheet Theory}
The basic idea of \cite{Berkovits:1993xq} is to find a matter CFT which reproduces the bosonic string. As the goal is to embed into the superstring, the ghost sector is the ghost sector of $\mathcal{N}=1$ superstring theory, i.e. the bc and $\beta\gamma$ ghosts. The matter sector can be any CFT with total central charge $c=15$ and $\mathcal{N}=1$ superconformal symmetry. They choose 26 free bosons $ X^i$ as well as a spin shifted bc-system with spin $(3/2,-1/2)$, which we will denote by\footnote{We use a different notation than Berkovits as we will use indices to describe the modes of the fields, such that $c_1$ will refer to the first mode of the $c$ ghost field instead of the spin shifted ghost, which is denoted $c^{\prime}$.} $b'c^{\prime}$. Note that this system is part of the matter sector and that the fields are fermions on the world-sheet with correct statistics. As usual the $\beta\gamma$ system is fermionized and replaced by an $\eta\xi$ system and a free boson $\phi$ with background charge $2$.

The total energy-momentum tensor is given by
\begin{equation}
T=T_X+T_{b^{\prime}c^{\prime}}+T_{bc}+T_{\beta\gamma}=T_X+T_{b^{\prime}c^{\prime}}+T_{bc}+T_{\xi\eta}+T_{\Phi}\;,
\end{equation}
with
\begin{align}
T_X=\sum_{i=1}^{26}\partial X_i\partial X^i\;,\\
T_{bc}=-2b\partial c-(\partial b)c\;,\\
T_{b^{\prime}c^{\prime}}=-{3\over 2}b^{\prime}\partial c^{\prime}-{1\over 2}(\partial b^{\prime})c^{\prime}+\partial^2(c^{\prime}\partial c^{\prime})\;,\\
T_{\beta\gamma}=-{3\over 2}\beta\partial \gamma-{1\over 2}(\partial \beta)\gamma\;,\\
T_{\eta\xi}=-\eta\partial\xi\;,\\
T_{\Phi}=-{1\over 2}\partial\Phi\partial\Phi-\partial^2\Phi\;.
\end{align}
Table \ref{t1} lists the conformal weights and Grassmann parities of all fields in the theory.
\begin{table}
\begin{center}
\vspace{0.15in}
\setlength{\arrayrulewidth}{.25mm}
\renewcommand{\arraystretch}{1.5}
\begin{tabular}{ |p{2cm}|p{3cm}|p{2.5cm}|p{2cm}|p{2cm}|p{1cm} | }
\hline
\bf{Field} & \bf{Conformal Weight $(\bar{h},h)$} & \bf{Grassmann Parity} & \bf{Ghost Number} & \bf{Picture Number} & \bf{GSO} \\
\hline
$\alpha=\partial X$&(0,1)&even&0&0&+\\
$\ov{\alpha}=\ov{\partial} \ov{X}$&(1,0)&even&0&0&+\\
\hline
$\beta$ & $(0,3/2)$ & even & -1 & 0 & - \\
$\gamma$ & $(0,-{1/2})$ & even & 1 & 0 & - \\
$\ov{\beta}$ & $(0,3/2)$ & even & -1 & 0 & - \\
$\ov{\gamma}$ & $(0,-{1/2})$ & even & 1 & 0 & - \\
\hline
$b$ & $(0,2)$ & odd & -1 & 0 & + \\
$c$ & $(0,-1)$ & odd & 1 & 0 & + \\
$\bar{b}$ & $(2,0)$ & odd & -1 & 0 & + \\
$\bar{c}$ & $(-1,0)$ & odd & 1 & 0 & + \\
\hline
$\eta$ & (0,1) & odd & 1 & -1 & + \\
$\xi$ & (0,0) & odd & -1 & 1 & +  \\
$\ov{\eta}$ & (1,0) & odd & 1 & -1 & + \\
$\ov{\xi}$ & (0,0) & odd & -1 & 1 & +  \\
\hline
$\phi$ & (0,1) & even & 0 & 0 & + \\
$e^{q\phi}$ & (0,$-q (q+2)/2$) & $(-1)^q$ & 0 & $q$ & $(-1)^q$ \\
$\ov{\phi}$ & (1,0) & even & 0 & 0 & + \\
$e^{q\ov{\phi}}$ & ($-q (q+2)/2$,0) & $(-1)^q$ & 0 & $q$ & $(-1)^q$ \\
\hline
\end{tabular}
\end{center}
\caption{The quantum numbers and conformal weights of various fields.
\label{t1}
} 
\vskip -.1in
\end{table}
The theory, in addition to the 26 bosons, consists of 4 first order systems with different weights and Grassmann parities. The description of these follows \cite{Blumenhagen:2013fgp}. Assume two conjugate fields b and c with energy momentum tensor
\begin{equation}
T=-\lambda b\partial c+(1-\lambda)(\partial b)c
\end{equation}
and define the background charge
\begin{equation}
Q=\epsilon(1-2\lambda)\;,
\end{equation}
where $\epsilon$ is the Grassmann parity of the system. Then the central charge of the system is 
\begin{equation}
c=\epsilon(1-3Q^2)\;.
\end{equation}
Moreover, the bc-current is non conserved:
\begin{equation}
N_c-N_b=\epsilon Q(g-1)=2\lambda-1\;,
\end{equation}
where we inserted $g=0$ for the sphere case we are interested in. Table \ref{t2} lists the attributes of the 4 first order systems of the theory. The fields are expanded into modes as
\begin{equation}
\mathcal{O}(z)=\sum_{n\in \mathbb{Z}}\mathcal{O}_nz^{-n-h}\;,
\end{equation}
for integer conformal weight and
\begin{equation}
\mathcal{O}(z)=\sum_{n\in \mathbb{Z}+1/2}\mathcal{O}_nz^{-n-h+1/2}\;,
\end{equation}
for half-integer weight in the NS sector. In the following we will only focus on the NS sector of the theory. The modes annihilate the Sl$(2,\mathbb{Z})$ vacuum $\ket{0}$ if the mode number is larger than the conformal weight:
\begin{equation}
\mathcal{O}_n\ket{0}=0\quad \text{if}\quad n>h\;.
\end{equation} 
Therefore in the NS sector all positive or zero-modes annihilate the vacuum except $c_1$, $c_0$, $c^{\prime}_{1/2}$ and $\xi_0$. The $\xi$ zero mode is not in the real Hilbert space as it is an artifact of the bosonization. Therefore, it is removed from the Hilbert space and implicitly added into all amplitudes to saturate the $\eta\xi$ current anomaly. The $c^+_0=c_0+\ov{c}_0$ mode is removed by Siegel gauge
\begin{equation}
b^+_0\ket{\Psi}=(b_0+\ov{b}_0)\ket{\Psi}=0\;,
\end{equation}
while the $c^-_0=c_0-\ov{c}_0$ mode is removed by acting with the BRST charge $Q_B$ on the level matching condition:
\begin{equation}
Q_B(L_0-\ov{L}_0)\ket{\Psi}=(b_0-\ov{b}_0)\ket{\Psi}=b_0^-\ket{\Psi}=0\;.
\end{equation}
The remaining positive modes lower the energy of the vacuum, so the true vacuum is not the $Sl(2,\mathbb{Z})$ vacuum $\ket{0}$ but 
\begin{equation}
\ket{0}_{NSNS}=c_1c^{\prime}_{1/2}\ov{c}_1\ov{c}^{\prime}_{1/2}\ket{0}\;.
\end{equation}


For a non-vanishing amplitude the current non-conservations have to be saturated. This requires 3 $c$~modes more then $b$~modes, 2 $c^{\prime}$~modes more than $b^{\prime}$~modes and 1 $\xi$~mode more than $\eta$~modes. The latter is already satisfied by the $\xi_0$~mode. These conditions hold for the holomorphic as well as the anti-holomorphic sector. Therefore the basic overlap is defined as
 \begin{equation}
 \label{basicoverlap}
 \bra{0}c_{-1}\ov{c}_{-1}c^{\prime}_{-1/2}\ov{c}^{\prime}_{-1/2}c_0^+c^-_0c_1\ov{c}_1c^{\prime}_{1/2}\ov{c}^{\prime}_{1/2}e^{-2\Phi-2\ov{\Phi}}\ket{0}=1\;,
 \end{equation}
or equivalently

\begin{equation}
 \bra{0}c_{-1}\ov{c}_{-1}c^{\prime}_{-1/2}\ov{c}^{\prime}_{-1/2}c_0\ov{c}_0c_1\ov{c}_1c^{\prime}_{1/2}\ov{c}^{\prime}_{1/2}e^{-2\Phi-2\ov{\Phi}}\ket{0}=2\;,
\end{equation}
which fixes the overall normalization.

\begin{table}
\begin{center}
\vspace{0.15in}
\setlength{\arrayrulewidth}{.25mm}
\renewcommand{\arraystretch}{1.5}
\begin{tabular}{ |p{2cm}|p{2.7cm}|p{2.3cm}|p{0.8cm}|p{2cm}|p{2cm}| }
\hline
\bf{System} & \bf{Conformal Dimension $\lambda$}&\bf{Grassmann Parity $\epsilon$} &$Q$& \bf{Central Charge $c$} & \bf{Current Violation}\\
\hline
$bc$&2&$1$&3&$-26$&3\\
$b^{\prime}c^{\prime}$&$3/2$&$1$&2&11&2\\
$\beta\gamma$&$3/2$&$-1$&$-2$&$-11$&2\\
$\eta\xi$&1&$1$&1&$-2$&$1$\\
\hline
\end{tabular}
\end{center}
\caption{The bc- and $\Phi$- systems of the theory.
\label{t2}
} 
\vskip -.1in
\end{table}

\subsection{Closed Superstring String Field Theory}

To write down the action we follow \cite{deLacroix:2017lif} and split the string field $\Psi$ into a set of string fields with definite picture numbers $n$ and $m$: 
\begin{equation}
\Psi=\bigoplus\limits_{n,m\in \mathbb{Z}}\Psi_{(n,m)}
\end{equation}
and defining 
\begin{align}
\Psi=\Psi_{(-1,-1)}\oplus\Psi_{(-1/2,-1)}\oplus\Psi_{(-1,-1/2)}\oplus\Psi_{(-1/2,-1/2)}\;,\\
\tilde{\Psi}=\Psi_{(-1,-1)}\oplus\Psi_{(-3/2,-1)}\oplus\Psi_{(-1,-3/2)}\oplus\Psi_{(-3/2,-3/2)}\;.
\end{align}
Furthermore, the operator $G$ is defined by acting with the zero mode of the picture changing operator on R states and doing nothing with NS states.
With these definitions the action can be written as \cite{deLacroix:2017lif}:
\begin{equation}
\label{action}
S=-{1\over 2}\bra{\tilde{\Psi}}c_0^+c_0^-L_0^+G\ket{\tilde{\Psi}}+\bra{\tilde{\Psi}}c_0^+c_0^-L_0^+\ket{\Psi}+\sum_{n=1}^\infty {1\over n!}\{\{\Psi^n\}\}\;.
\end{equation}

In this paper we will only consider terms up to quartic order and the genus $0$ contribution. Therefore the action becomes
\begin{equation}
S=-{1\over 2}\bra{\tilde{\Psi}}c_0^+c_0^-L_0^+G\ket{\tilde{\Psi}}+\bra{\tilde{\Psi}}c_0^+c_0^-L_0^+\ket{\Psi}+{\{\{\Psi^3\}\}\over 6}+{\{\{\Psi^4\}\}\over 24}\;,
\end{equation}
where
\begin{equation}
{\{\{\Psi^3\}\}}=\{f_1\circ\Psi(0),f_2\circ\Psi(0),f_3\circ\ov{X}X\Psi(0)\}
\end{equation}
and
\begin{equation}
{\{\{\Psi^4\}\}}=\int_{V_{0,4}} d\xi d\overline{\xi}\bra{\Sigma} BB^{\star}\ket{\Psi_1}\ket{\Psi_2}\ket{X\ov{X}\Psi_3}\ket{X\ov{X}\Psi_4}\;.
\end{equation}
Technical details on the used conformal transformations $f_i$, the chosen picture changing conventions as well as how the n-point functions are evaluated can be found in the appendix. In this paper we are focusing on the NSNS sector up to quartic order, where the action takes the form
\begin{equation}
\label{potential}
S={1\over 2}\bra{\Psi}c_0^+c_0^-L_0^+\ket{\Psi}+{\{\{\Psi^3\}\}\over 6}+{\{\{\Psi^4\}\}\over 24}\;.
\end{equation}

The string field is expanded in the level $l=L_0-2$. In this paper we only investigate up to level 2, i.e. massless fields.

We now turn to the construction of the string field $\Psi$. The left and right moving sectors are independent of each other up to the level matching condition
\begin{equation}
L_0=\overline{L_0}\;.
\end{equation}
Therefore the string field is constructed by combining open string states level by level. The only condition imposed is picture $-1$ in the NS-sector and picture $-1/2$ in the R-sector. To effectively generate these states the generating function
\begin{align*}
f(x,z,\ov{x},\ov{z})=\prod_{i=1}^{N_{bos}}\prod_{j=h(i)}^{\infty}{1\over 1-\mathcal{O}_{-j}^{(i)}x^{j}z^{pic(i)}}\\\times\prod_{k=1}^{N_{fer}}\prod_{j=h(k)}^{\infty}(1-\mathcal{O}^{(k)}_{-j}x^{j}z^{pic(k)})
\end{align*}
is used, where $pic(i)$ is the picture number of the i-th operator and expanding it in $x$ and $z$. The exponents of these then give the conformal weight and picture number of the states. $\mathcal{O}$ are all fields of the theory, i.e. $b$, $c$, and $\alpha$ for the bosonic string and additionally $b'$, $c^{\prime}$, $\eta$, $\xi$ and $\phi$ in the superstring case.


\section{The Action}
\subsection{Embedding of the Bosonic SFT}
Before we evaluate \eqref{potential} for the $\mathcal{N}=1$ CSFT, we recall first bosonic CSFT. In bosonic CSFT and Siegel gauge the string field up to level $l=2$ reads
\begin{align*}
\psi_{bos}=t(x)c_1\ov{c}_1\ket{0}
+d_1(x)c_{-1}c_1\ov{c}_{-1}\ov{c}_1\ket{0}+d_2(x)c_{-1}c_1&\ket{0}\\+d_3(x)\ov{c}_{-1}\ov{c}_1\ket{0}
+g_{\mu\nu}(x)\alpha^{\mu}_{-1}c_1\ov{\alpha}^{\nu}_{-1}\ov{c}_1&\ket{0}\\
+A_{\mu,1}(x)\alpha^{\mu}_{-1}c_1\ov{c}_{-1}\ov{c}_1\ket{0}+A_{\mu,2}(x)c_{-1}c_1\ov{c}_{1}\ov{\alpha}^{\mu}_{-1}&\ket{0}\\
+A_{\mu,3}(x)\alpha^{\mu}_{-1}c_1\ket{0}+A_{\mu,4}(x)\ov{c}_{1}\ov{\alpha}^{\mu}_{-1}&\ket{0}\\
+I(x)&\ket{0}\;.
\end{align*}
These are in total 10 fields, the tachyon $t$, 3 dilaton scalars $d_i$, 4 vectors $A_{\mu}$, the graviton $g_{\mu\nu}$ and the Sl$(2,\mathbb{C})$vacuum denoted $I$. Evaluating the action for this string field results in 
\begin{equation}
\label{bospotential}
V(\psi)=-t^2+{27\over 32}td_2d_3+{6561\over 4096}t^3+{27\over 32}Itd_1+{27\over 16}g_{\mu\nu}g^{\mu\nu}t\;.
\end{equation}
This is the full closed bosonic string field potential up to qubic order in Siegel gauge. Normally if one is only interested in the tachyon potential the first 3 terms are taken into account, as the latter have no effect on the tachyon potential due to twist symmetry. But they provide a nontrivial check that the embedding is working. For the supersymmetric case the evaluation of the action has to be automated. The fact that these terms get reproduced serves as a validation of the code. Note also that the vectors do not appear at the cubic level.\\
Focus for a moment on the second term in \eqref{bospotential}. With the field redefinitions
\begin{align}
\label{redef}
\ket{d}\;=\ket{d_2}-\ket{d_3}=(c_{-1}c_1-\ov{c}_{-1}\ov{c}_1)&\ket{0}\\
\ket{d_g}=\ket{d_2}+\ket{d_3}=(c_{-1}c_1+\ov{c}_{-1}\ov{c}_1)&\ket{0}\;,
\end{align}
it takes the form 
\begin{equation}
-{27\over 128}td^2+{27\over 128}td_g^2\;.
\end{equation}
where $d$ now is the ghost dilaton and $d_g$ is pure gauge.\footnote{This is not equal to the standard form of the potential which is $-{27\over 32}td^2$. The reason is that we have included $d_g$, by choosing $d_g^2=-3d^2$ the action takes the standard form. While this condition seems rather strange, it is the result of setting $d_g=0$ in the string field. The factor of 4 can be seen by the simple calculation $d_2\cdot d_3={d^2-d_g^2\over4}={d^2\over 4}$ upon setting $d_g=0$.} We will mostly work with the fields $d_2$ and $d_3$ and their analogs in the $\mathcal{N}=1$ case, but it is important to keep in mind that we still have to fix the gauge in the end. Also note that a non-zero vacuum expectation value(VEV) for the tachyon results in a symmetric splitting of the masses in the effective field theory, one becoming tachyonic and one massive. Of course the contributions of higher level field counter this effect such that the dilaton remains a marginal direction \cite{Yang:2005ep}.\\
The $\mathcal{N}=1$  counter parts of these fields are obtained by multiplying them with $e^{-\phi-\ov{\phi}}c^{\prime}_{1/2}\ov{c}'_{1/2}$. \eqref{bospotential} gets then reproduced by the fact that the contributions from $e^{-\phi}$ and $c^{\prime}_{1/2}$ exactly cancel each other and that the only term contributing from the PCO is $e^{\phi}b'$ due to $c^{\prime}b'$ conservation, such that 
\begin{equation}
\ov{X}Xe^{-\phi-\ov{\phi}}c^{\prime}_{1/2}\ov{c}'_{1/2}V \approx V\;,
\end{equation}
where V is built out of $c$, $b$ and $\alpha$ modes. Moreover,
\begin{equation}
\Braket{e^{-\phi-\ov{\phi}}c^{\prime}_{1/2}\ov{c}'_{1/2}V_1'\;,\;e^{-\phi-\ov{\phi}}c^{\prime}_{1/2}\ov{c}'_{1/2}V_2\;,\;V_3}_{\mathcal{N}=1}=\Braket{V_1,V_2,V_3}_{bos}\;.
\end{equation}
Therefore, the bosonic moduli space is a subspace of the superstring moduli space. We now turn to the superstring case.
\subsection{The $\mathcal{N}=1$ String Field}

In the NSNS sector the string field of the $\mathcal{N}=1$ theory consists sector out of the following fields. At level zero there is only the tachyon
\begin{equation}
t=t(x)\cdot c_1\ov{c}_1c^{\prime}_{1/2}\ov{c}^{\prime}_{1/2}e^{-\Phi-\ov{\Phi}}\ket{0}\;.
\end{equation}
At level 1 there are 16 additional tachyons which will be labeled $B_i$, i=1..16. They are listed in table \ref{tachyonlist}. We will refer to them as "secondary" tachyons.  At level 2 there are 121 additional massless fields, 9 of which correspond to fields which are also present in the bosonic SFT. We label these 9 as in the bosonic case.\\
\renewcommand{\arraystretch}{1.3}
\begin{table}
\begin{center}
\begin{tabular}{|l|l|}
\hline
$t$&$c_1 \bar{c}_1 e^{-\bar{\phi }-\phi } c^{\prime}_{\frac{1}{2}} \overline{c^{\prime}}_{\frac{1}{2}}
   $\\[6pt]
\hline
 $B_1$ & $c_1 \bar{c}_1 e^{-\bar{\phi }-\phi } $\\[6pt]
 \hline
 $B_2$ & $c_1 \bar{c}_1 e^{-\bar{\phi }-\phi } c^{\prime}_{-\frac{1}{2}} c^{\prime}_{\frac{1}{2}} $\\[6pt]
 \hline
$ B_3$ & $c_1 \eta _{-1} \bar{c}_1 e^{-\bar{\phi }} c^{\prime}_{\frac{1}{2}} $\\[6pt]
 \hline
 $B_4$ & $c_1 \xi _{-1} \bar{c}_1 e^{-\bar{\phi }-2 \phi } c^{\prime}_{\frac{1}{2}} $\\[6pt]
 \hline
$ B_5$ & $c_1 \bar{c}_1 e^{-\bar{\phi }-\phi } \overline{c^{\prime}}_{-\frac{1}{2}} \overline{c^{\prime}}_{\frac{1}{2}} $\\[6pt]
 \hline
 $B_6$ & $c_1 \bar{c}_1 e^{-\bar{\phi }-\phi } c^{\prime}_{-\frac{1}{2}} c^{\prime}_{\frac{1}{2}} \overline{c^{\prime}}_{-\frac{1}{2}}
  \overline{c^{\prime}}_{\frac{1}{2}} $\\[6pt]
  \hline
 $B_7$ & $c_1 \eta _{-1} \bar{c}_1 e^{-\bar{\phi }} c^{\prime}_{\frac{1}{2}} \overline{c^{\prime}}_{-\frac{1}{2}} \overline{c^{\prime}}_{\frac{1}{2}} $\\[6pt]
 \hline
 $B_8$ & $c_1 \xi _{-1} \bar{c}_1 e^{-\bar{\phi }-2 \phi } c^{\prime}_{\frac{1}{2}} \overline{c^{\prime}}_{-\frac{1}{2}} \overline{c^{\prime}}_{\frac{1}{2}} $\\[6pt]
 \hline
$ B_9$ & $c_1 e^{-\phi } \bar{c}_1 \bar{\eta }_{-1} \overline{c^{\prime}}_{\frac{1}{2}} $\\[6pt]
 \hline
 $B_{10}$ & $c_1 e^{-\phi } \bar{c}_1 \bar{\eta }_{-1} c^{\prime}_{-\frac{1}{2}} c^{\prime}_{\frac{1}{2}} \overline{c^{\prime}}_{\frac{1}{2}} $\\[6pt]
 \hline
$ B_{11}$ & $c_1 \eta _{-1} \bar{c}_1 \bar{\eta }_{-1} c^{\prime}_{\frac{1}{2}} \overline{c^{\prime}}_{\frac{1}{2}} $\\[6pt]
 \hline
 $B_{12}$ & $c_1 \xi _{-1} e^{-2 \phi } \bar{c}_1 \bar{\eta }_{-1} c^{\prime}_{\frac{1}{2}} \overline{c^{\prime}}_{\frac{1}{2}} $\\\hline
 $B_{13}$ & $c_1 \bar{c}_1 \bar{\xi }_{-1} e^{-2 \bar{\phi }-\phi } \overline{c^{\prime}}_{\frac{1}{2}} $\\[6pt]
 \hline
$ B_{14}$ & $c_1 \bar{c}_1 \bar{\xi }_{-1} e^{-2 \bar{\phi }-\phi } c^{\prime}_{-\frac{1}{2}} c^{\prime}_{\frac{1}{2}} \overline{c^{\prime}}_{\frac{1}{2}} $\\[6pt]
\hline
$ B_{15}$ & $c_1 \eta _{-1} \bar{c}_1 \bar{\xi }_{-1} e^{-2 \bar{\phi }} c^{\prime}_{\frac{1}{2}} \overline{c^{\prime}}_{\frac{1}{2}} $\\[6pt]
 \hline
 $B_{16}$ & $c_1 \xi _{-1} \bar{c}_1 \bar{\xi }_{-1} e^{-2 \bar{\phi }-2 \phi } c^{\prime}_{\frac{1}{2}} \overline{c^{\prime}}_{\!\frac{1}{2}} $\\[6pt]
 \hline
\end{tabular}
\end{center}
 \caption{List of tachyonic states of the theory. t has weight $-2$, all $B_i$ have weight $-1$.}
 \label{tachyonlist}
\end{table}

\subsection{The Kinetic Terms}
In the NS-sector the kinetic part of \eqref{action} simplifies to
\begin{equation}
S_{kin,NS}={1\over 2}\bra{\Psi}c^+c^-L_0\ket{\Psi}={h_{\Psi}\over 2}\bra{\Psi}c^+c^-\ket{\Psi}\;.
\end{equation}
The brackets can be evaluated using the BPZ inner product. For the tachyon $t=t(x)\cdot c_1\ov{c}_1c^{\prime}_{1/2}\ov{c}^{\prime}_{1/2}e^{-\Phi-\ov{\Phi}}\ket{0}$ with $h_t=-2$ this gives using the basic overlap \eqref{basicoverlap}
\begin{equation}
\begin{split}
-\bra{t}c^+c^-\ket{t}&=-t^2\bra{0}c_{-1}\ov{c}_{-1}c^{\prime}_{-1/2}\,\ov{c}^{\prime}_{-1/2}c_0^+c^-_0c_1\ov{c}_1c^{\prime}_{1/2}\ov{c}^{\prime}_{1/2}\;e^{-2\Phi-2\ov{\Phi}}\ket{0}\\&=-t^2\;.
\end{split}
\end{equation}
The kinetic terms for the massless level vanish due to $L_0\ket{\psi}=0$. The secondary tachyons all have $h=-1$. The different ghost number conservations cause most of the possible quadratic terms to vanish. Only 4 terms survive, which after taking the BPZ-conjugate and resorting have exactly the form of the basic overlap. For the surviving terms $\bra{a}c^+c^-\ket{b}=\bra{b}c^+c^-\ket{a}$, which gives another factor of $2$, such that the total quadratic potential becomes
\begin{equation}
\label{A2}
V_2^{(2)}=-t^2+B_2B_5-B_4B_7+B_{10}B_{13}+B_{11}B_{16}\;.
\end{equation}
Going to an orthogonal mass basis, one can see that there are actually only 4 additional tachyons, 4 massive fields and 8 massless ones. 
\subsection{The Cubic Terms}
After imposing ghost number constraints, there are 3268 terms to evaluate. Explicit calculation of them shows that there are 972 non-vanishing 3-vertices. In this section we will discuss the structure of the resulting action  and its equations of motion. 
The complete action is too complicated to reproduce it here in full detail.\footnote{The full action can be found in the source files, see appendix \ref{data}.} We only present the tachyonic part. 

\begin{align*}
V_1^{(3)}=&-\frac{243 B_1^2 t}{256}+\frac{243}{128} B_1 B_2 t+\frac{243}{128} B_1 B_5 t-\frac{243}{128} B_2 B_5 t\\&-\frac{1215}{512} B_1 B_6 t-\frac{243}{128} B_{12} B_{15} t+\frac{243}{128} B_{11} B_{16} t+\frac{6561 t^3}{4096}
\end{align*}
If one includes the counterparts of the $9$ massless fields of bosonic SFT, one recovers \eqref{bospotential} without any additional couplings between the new tachyons $B_i$ and the bosonic fields. Also including all massless fields, there is no direct coupling between the bosonic fields and the additional tachyons. But there is an indirect coupling because of the additional massless fields coupling to the tachyons in various constellations as well as to the bosonic fields.

Out of the $138$ fields, $127$ appear linearly in the action. But this is mainly an artifact of the chosen basis for the string field. Field redefinitions similar to \eqref{redef} can significantly reduce this number. 
\subsection{The Quartic Terms}
The evaluation of the action at the quartic level up to the massless level requires the evaluation of $25\cdot 64^2\cdot 138^4=3.7\cdot 10^{13}$ terms, where the 25 comes from the possibly contributing b-ghost insertions, the $64^2$ from the two PCO insertions and the $138^4$ from the 4 string fields. Even after imposing ghost number conservation over 4.2 million quartic vertices remain. For each of these one has to use the conservation laws and then integrate over the moduli space. Unfortunately, our code is not powerful enough to handle this amount of terms, especially the numerical integration to high enough precision takes between $2$ and $6$ seconds, resulting in a time consuming calculation.

Instead we calculated all quartic terms at level 1, with the result

\begin{align*}
V_1^{(4)}=&-3.02 t^4\\
&-t^2\Bigl(12.88 B_1^2-11.37 B_2 B_1-36.62 B_5 B_1+75.59 B_6 B_1+12.81 B_2 B_5\\&\qquad\;+52.64 B_4 B_7-52.64 B_3 B_8+7.00 B_{10} B_{13}-7.00 B_9 B_{14}\\&\qquad\;+386.48 B_{12} B_{15}-386.48 B_{11} B_{16}\Bigr)\\&+9943.20 B_1^2 B_6^2\;.
\end{align*}

The calculation of these terms is presented in appendix \ref{terms}. For each qubic term $tB_iB_j$ there is a corresponding $t^2B_iB_j$ term. Additionally, there are new terms coupling the remaining secondary tachyons quadratically to the primary tachyon. There is only a single quartic state not coupling to the primary tachyon.
We will study the effects of these terms in the next section. In \cite{Yang:2005ep} many bosonic quartic terms were calculated, but as we are unable to calculate all massless quartic terms we will not include them to avoid a bias towards the bosonic fields.

\section{Solving the Equations of Motion}
The tachyon couples to all fields at least once, such that in principle one has to include all fields for an exact solution. As there are 138 fields, this is a tremendous task. Instead, we will start by looking only at the level $0$ and $1$ fields, solve those equations and see how many of these solutions survive at the higher levels.

\subsection{Level 0}
At level $l=0$ there is only the tachyon with the full potential
\begin{equation}
V_0=-t^2+{6561\over 4096}t^3-3.01625t^4+\mathcal{O}(t^5)\;.
\end{equation}
At qubic order, this has two solutions. First the perturbative vacuum at t=0 and second a minimum at $t={8192\over19683}\approx0.416$. This minimum gets destroyed by the quartic term but restored by the quintic one.

\subsection{Level 1}
At level 1 there are 16 more fields to take into account. 8 of these have a kinetic term, see \eqref{A2}. Diagonalising these, 4 are tachyons of weight $-1$ and 4 massive fields of weight $1$. 
\subsubsection*{Cubic Order}
We start at cubic order. The four fields $B_3$, $B_8$, $B_9$ and $B_{14}$ do not appear at this order. Therefore there are 13 equations to solve. 4 of these simply set the tachyons which appear only in their kinetic terms to 0, i.e. $B_4$, $B_7$, $B_{10}$ and $B_{13}$. 
The remaining 9 EOMs have 6 solutions. These are
\begin{enumerate}
\item The vacuum solution, $t=0, B_i=0$\;.
\item The minimum of the bosonic theory, $t=0.416$, $B_i=0$\;.
\item $t=0$, $B_6=-\frac{2 B_1}{5}$\;.
\item $t=0$, $B_{15}=\frac{-2 B_1^2-5 B_6 B_1}{4 B_{12}}$\;.
\item $t=+{128\over 243}$, $B_5= \frac{8704}{59049 B_2}$,$B_6= \frac{4 \left(59049 B_2^2+8704\right)}{295245 B_2}$\;.
\item $t=-{128\over 243}$, $B_{16}= -\frac{74240}{59049 B_{11}}$\;.
\end{enumerate}

All these solutions are only saddle points, i.e. all first derivatives of the fields vanish but there are remaining tachyons. There is no minimum at this level. The value of the potential is $0$ for the  $t=0$ solutions, $-0.0577399$ for the bosonic minimum, $-0.511575$ for the $t=-{128\over 243}$ solution and $-0.0433538$ for the $t=+{128\over 243}$ solution. All $t=0$ solutions have the same spectrum and value of the potential. It is likely that these are just gauge copies of each other. The masses of the fields are shown in table \ref{table4}. All solutions still contain tachyons. There is only one solution which is 'less' tachyonic than the vacuum solution. 
\begin{center}
\begin{table}[h]
\begin{scriptsize}
\begin{tabular}{|c|c|c|c|c|c|c|c|c|c|c|c|c|c|c|}
\hline
t&V&m&m&m&m&m&m&m&m&m&m&m&m&m\\
\hline
0&0& -2 & -1 & -1 & -1 & -1  & 0 & 0 & 0 & 0 & 1 & 1 & 1 & 1 \\\hline
0.42&-0.058& -$\frac{113}{16}$ & -2 & -1.09 & -1 & -1 & -1 & 0  & 0 & 0.91 & 1 & 1 & 1 & 3.18 \\\hline
$+\frac{128}{243}$&-0.04&-2.18 & -2 & -1 & -1 & -1   & 0 & 0 & 1 & 1 & 1 & 1.18 & 2 & $\frac{49}{16}$ \\\hline
$-\frac{128}{243}$&-0.51&-1.90 & $-\frac{145}{81}$ & -1 & -1 & $-\frac{64}{81}$ & $-\frac{17}{81}$  & 0.09 & $\frac{64}{81}$ & 1 & 1 & 1.23 & $\frac{145}{81} $& 2 \\\hline
\end{tabular}
\caption{The values of the tachyon, the potential and the masses of the fields in the 4 different solutions at cubic order. This excludes the 4 massless fields which do not appear at cubic order.}
\label{table4}
\end{scriptsize}
\end{table}
\end{center}
\subsubsection*{Quartic Order}
\begin{figure}[ht]
  \centering
  \includegraphics[width=1\textwidth]{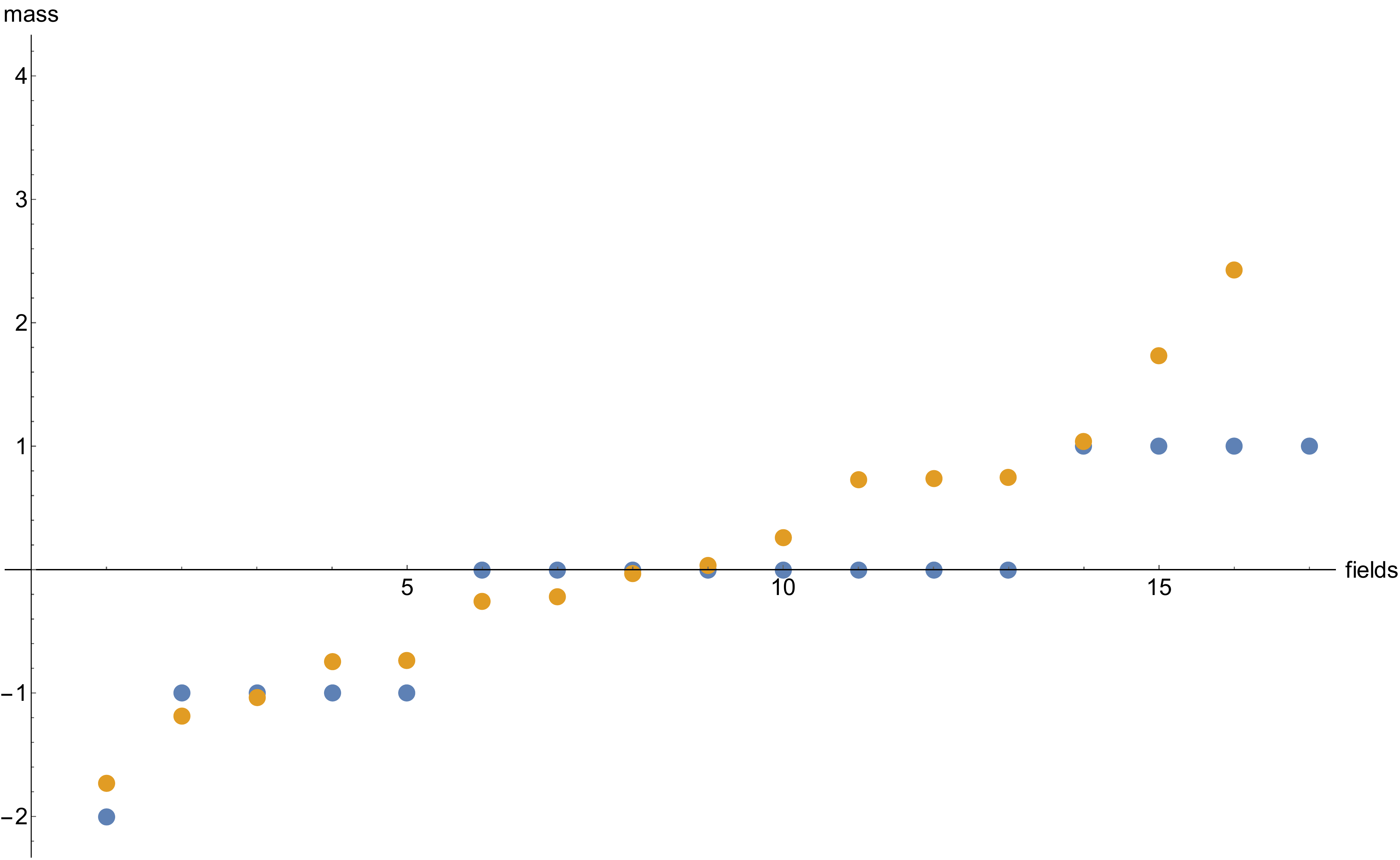}
  \caption{The masses of the fields in the vacuum (blue) and the other solution (orange). The last mass in the orange solution is extremely heavy with a mass of $\approx 2128$.}
  \label{masses}
\end{figure}

Now we turn to the quartic terms. If these are included all fields appear in the action, such that now there are 17 equations to solve. Also there are no longer equations which force fields to 0. The EOMs have in total 21 different solutions. If one imposes the reality constraint on the fields, i.e. one discards imaginary solutions, and counts gauge copies only once 7 different solutions survive. Most of these are highly tachyonic and seem to be numerical artifacts, the only interesting solutions are the vacuum $t=0$ and 
\begin{equation}
\label{solution}
t=0.069518\quad B_1=0.32693\quad B_2=0.015816 \quad B_5=-0.027077\quad B_6=-0.00003\;,
\end{equation}
while all other fields are 0. The bosonic minimum no longer appears in the solutions. But this also happens in the bosonic theory, where it only reappears at high level  \cite{Moeller:2006cv}. 
Figure \ref{masses} shows the masses of solution \eqref{solution} relative to the vacuum solution. There are no longer any massless fields. The lowest lying mass with $\approx-1.74$ is less tachyonic then in the vacuum. That this value is no longer $-2$ makes the interpretation of this solution very hard, as there is no longer a candidate for the type II tachyon. Including the fields of bosonic SFT, the trace of the metric and the ghost-dilaton obtain a small but non-zero mass at the the solution \eqref{solution}. This violates the marginality of the dilaton deformations and is clearly not the case in a flat space vacuum.

At the solution \eqref{solution} four secondary tachyons have a fixed value. We define a direction in the string field space by 
\begin{equation}
x= 0.32693 B_1+0.015816 B_2-0.027077 B_5-0.00003 B_3
\end{equation}
and integrate out all fields except $x$ and $t$. The quartic non vacuum solutions then lie at $x =\pm 1$. The effective potential is
\begin{equation}
V_{eff}=-3.020 t^4+\frac{6561 t^3}{4096}+1.640 x^2 t^2-t^2-0.108 x^2 t+ 10^{-6} x^4-0.0004 x^2\;.
\end{equation}

\begin{figure}[ht]
  \centering
  \includegraphics[width=1\textwidth]{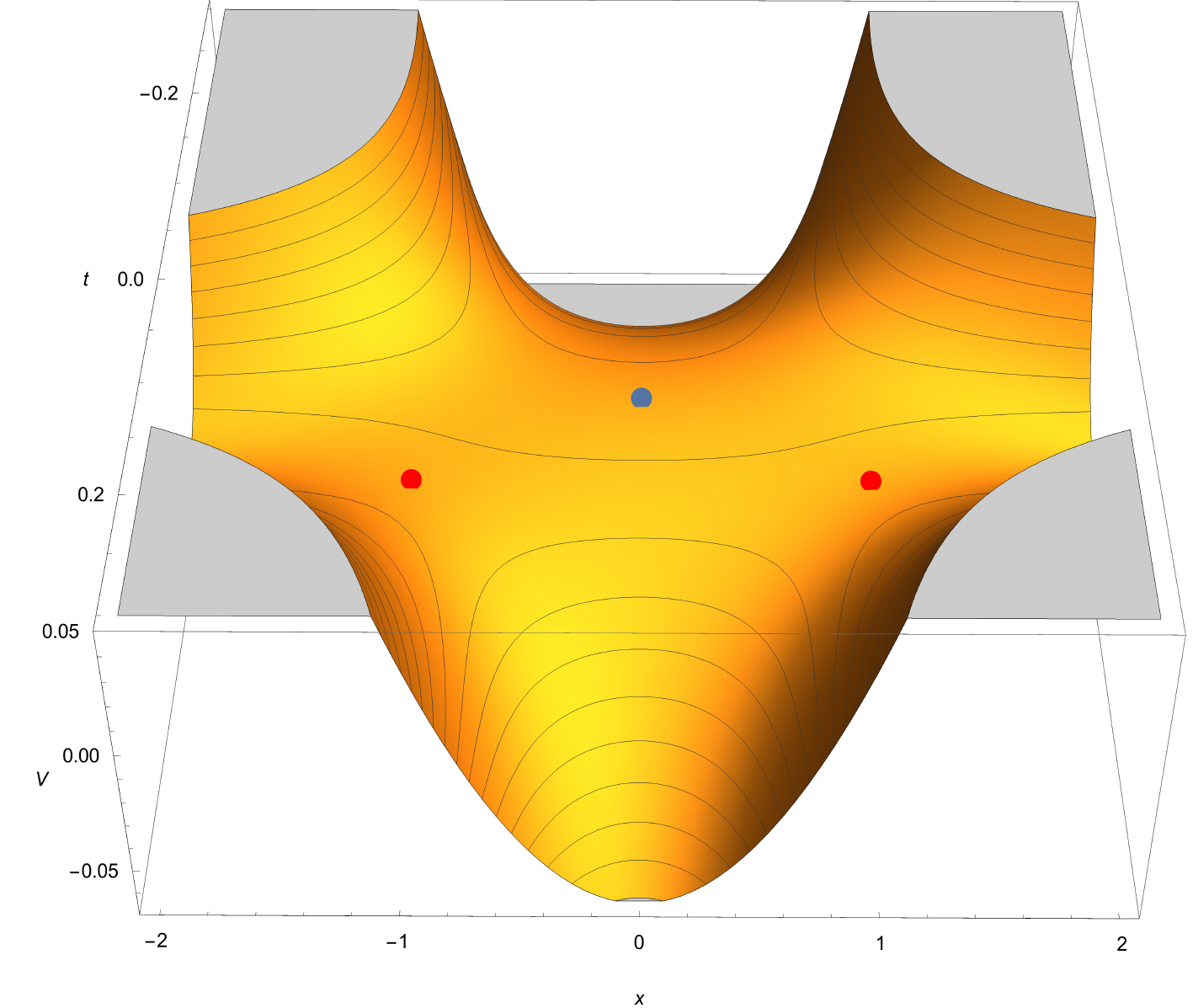}
  \caption{Effective quartic potential $V(t,x)$. The blue dot is the positions of the vacuum, the red dots are the other solutions at $x=\pm 1$. All solutions are only saddle points and show run-away behavior to infinity. The symmetry in x direction originates from unfixed twist symmetry.}
  \label{plt4}
\end{figure}

\begin{figure}[ht]
  \centering
  \includegraphics[width=1\textwidth]{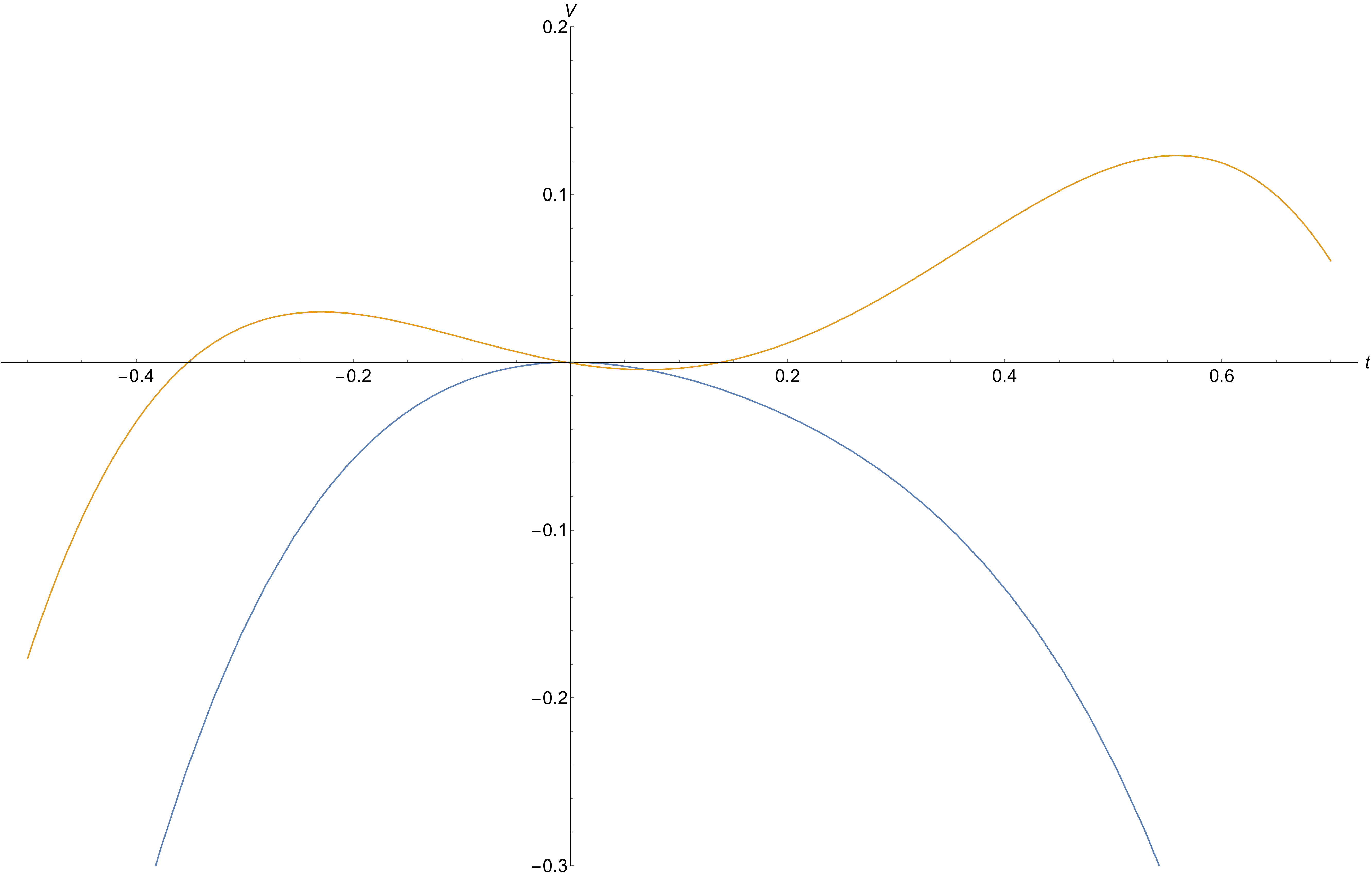}
  \caption{Effective quartic potential $V(t)$ for the vacuum solution(blue) and solution \eqref{solution} (orange). There is a minimum in the tachyon potential in the latter case despite the quartic contact term.}
  \label{plt5}
\end{figure}

Figure \ref{plt4} shows this effective potential. The symmetry in the x direction is obvious, as the effective potential contains only even powers of x. Figure \ref{plt5} shows the effective potential when all fields except the tachyon are integrated out at the two solutions. From these pictures it is clear that the effective potentials give a wrong picture about the physics and that one has to take all fields into account. In the pure tachyon potential it seems as if there is a minimum, but there are other tachyonic fields which render the solution only a saddle point. 

\begin{figure}[ht]
  \centering
  \includegraphics[width=1\textwidth]{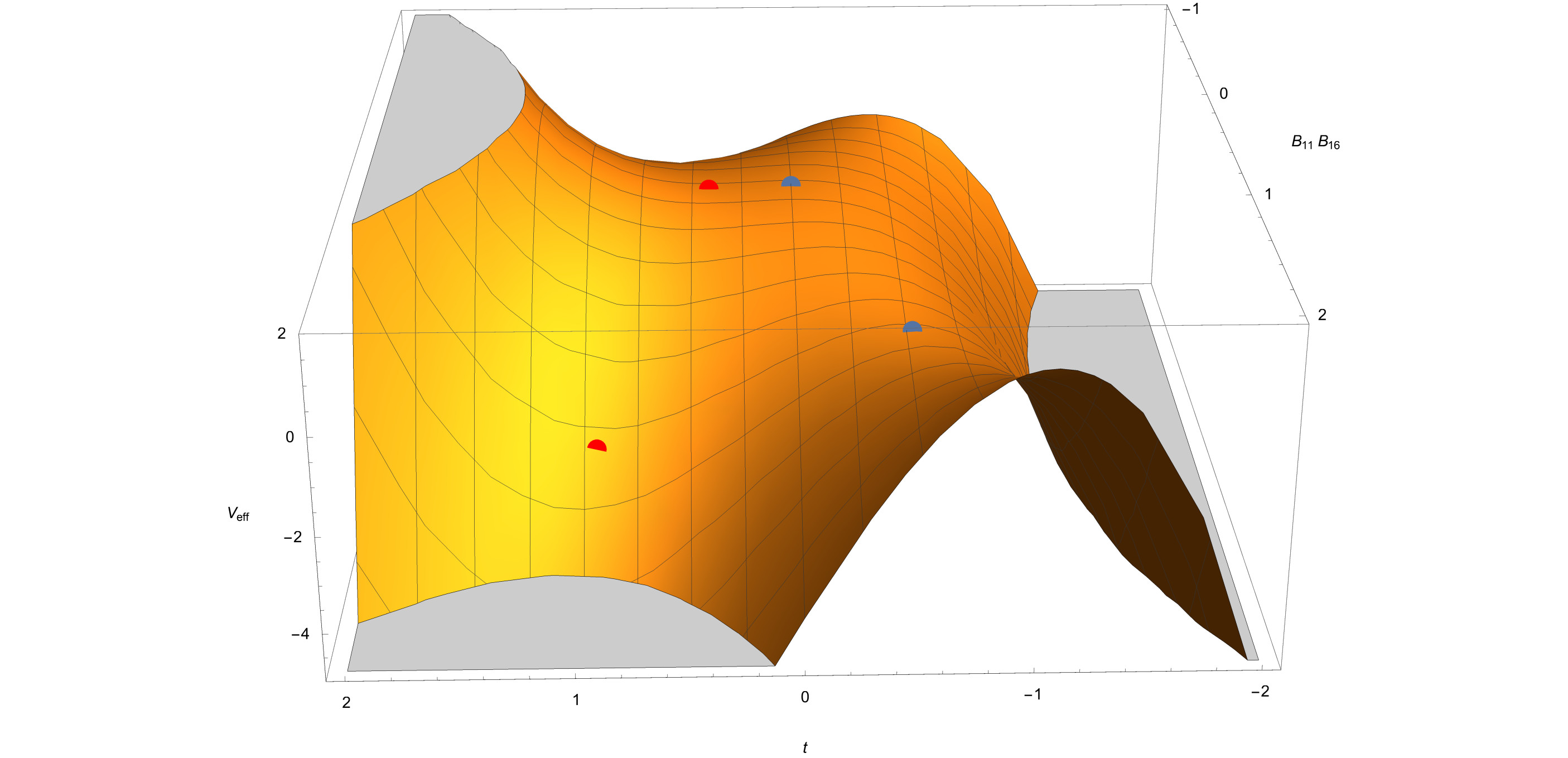}
  \caption{Effective potential $V(t)$ for a sample solution. The blue dots are the positions of the  solutions, the red dots are the minima of the 1-dimensional effective potentials at the solutions. They are destabilized and show run-away behavior to infinity.}
  \label{plt2}
\end{figure}

The runaway behavior is an artifact of only taking terms up to quartic order into account. This can be seen by comparing this effective potential to the corresponding one at cubic order. We choose the last solution of the cubic tachyonic system as an example. Integrating out all fields except $B_{11}$ and $B_{16}$, one obtains an effective two-dimensional potential, which is shown in figure \ref{plt2}. The form of the potential looks quite similar compared to the cubic case. There is a rather flat direction between the solutions, which we denoted $x$ in the quartic case and $B_{11}B_{16}$ in the cubic case. In the tachyon direction the potential is rather steep. The runaway directions appear in the region where the $t^3$ or $t^4$ terms dominate the potential. This is a strong hint that these are just artifacts of the expansion to cubic or quartic order.

\newpage

\subsection{Level 2}
At this level there are 138 fields, which all appear in the action. We were not able to find all solutions at this level. Simply setting all massless fields to zero is only a solution to the full set of EOMs for the vacuum solution, in the cubic as well as in the quartic case. Following the ideas of  \cite{Kudrna:2018mxa} and taking the level 1 solutions as a starting point for a Newton-Raphson method leads to a run-away behavior for all solutions, again for the cubic as well as the quartic case. Including only the massless fields of bosonic string field theory does not change the solutions, it just produces constraints on the additional fields. E.g. for the non-vacuum solution the constraint is setting all additional fields to 0. 
Moreover, the tachyons couple the massless fields in a complicated way, such that there are no subsets of fields only interacting among themselves. Thus there is no natural way of selecting additional fields to take into account. Especially since all fields couple in some way to the primary tachyon.

While we were not able to find a general solution with non-zero tachyon, there is no reason why there should not be one. The EOMs are a coupled set of cubic equations, so one would rather expect more solutions at a higher level. The issue in solving them lies in the runaway behavior of the potential, complicating a numerical solution. 138 equations are too many to be solved by a homotopy deformation method. This complication is also the reason why we did not try to evaluate the action at higher levels, as solving the EOMs would be even more cumbersome. 
\newpage
\section{Conclusions}
All found solutions are only saddle points of the potential, they still contain tachyons. This instability leads to potentials unbounded from below. This includes the minimum found in bosonic string field theory, which due to additional directions no longer is a minimum. Inclusion of the quartic tachyon terms does not reintroduce this minimum. While it could exist at the massless level, the runaway behavior of the Newton-Raphson method starting at this point in moduli space is a hint that it is no solution in the supersymmetric case. We were not able to find a new stable minimum, which could have been a candidate for a transition to superstring theory. Hopefully, the inclusion of the Ramond sector, higher levels of the string field and loop amplitudes could change this. Especially the lack of a candidate for the type II tachyon as well as the massive dilatons seem problematic. But assuming the endpoint of the tachyon condensation is not a type II theory, these would also be advantageous features, as e.g. couplings to the graviton as well as graviton VEVs are needed for spontaneous compactification.

The amount of secondary tachyonic fields obtaining a non-zero VEV increases at quartic order. Inclusion of quintic and higher terms could well give a VEV to all 16 secondary tachyons and produce a stable vacuum. But these computations are complicated due to the integration over the moduli space of n-punctured spheres. Already for the quintic terms the moduli space becomes too complicated to be represented as fits \cite{Moeller:2006cw}. Another direction would be an increase of the string field level. The enormous efforts required to reach higher levels  can be seen in the more advanced calculations in open string field theory, e.g. in \cite{Kudrna:2018mxa}. But the increased amount of fields also complicates solving the EOMs. All calculations were only preformed at tree level, i.e. the sphere case. There is at the moment no prescription available how to include the torus contribution to the action. Without a new method to evaluate the potentials or an analytic solution it will be difficult to obtain conclusive results. Unfortunately, despite some recent progress in the calculation of minimal area metrics \cite{Headrick:2018dlw,Headrick:2018ncs,Naseer:2019zau} or the proposed use of hyperbolic metrics \cite{Moosavian:2017fta,Moosavian:2017sev,Moosavian:2017qsp}, no method known to the author is developed far enough to conduct this calculation.

Berkovits and Vafa did also show that the $\mathcal{N}=1$ moduli space can be embedded into the $\mathcal{N}=2$ theory \cite{Berkovits:1993xq}. The critical points of this theory would be interesting to investigate. It is known that the tachyon condensation from the classical $\mathcal{N}=2$ superstring in a linear dilaton background leads to the bosonic string \cite{Hellerman:2007ym}. Therefore, the $\mathcal{N}=2$ string should have a higher value of the potential than the bosonic string. If there are critical points below the bosonic string, the interpretation of these is no longer clear. 

\noindent
\subsubsection*{Acknowledgments}
I like to thank Ralph Blumenhagen for very helpful discussions and comments about the manuscript. I also like to thank Dimitris Skliros for enlightening discussions about the geometry of superstring theory and string field theory.

\appendix
\section{Technical Details}
In this appendix we will discuss the technical details of the evaluation of the action, taking special care of signs and factors of 2. Especially the signs are sometimes subtle but have a significant effect on the result.
\subsection{Evaluating n-point Vertices}
As stated in section 2 we normalize the basic overlap such that the tachyon kinetic term is $-t^2$, i.e.

\begin{equation}
 \bra{0}c_{-1}\ov{c}_{-1}c^{\prime}_{-1/2}\ov{c}^{\prime}_{-1/2}c_0\ov{c}_0c_1\ov{c}_1c^{\prime}_{1/2}\ov{c}^{\prime}_{1/2}e^{-2\Phi-2\ov{\Phi}}\ket{0}=2\;.
\end{equation}

If one wants to split the closed string amplitudes into open string amplitudes, one has to careful take into account the ordering. E.g. for the bosonic case one has  \cite{Yang:2005iua}:
\begin{equation}
\bra{0}c_1^{(1)}c_1^{(2)}c_1^{(3)}\ov{c}_1^{(1)}\ov{c}_1^{(2)}\ov{c}_1^{(3)}\ket{0}=-2\bra{0}c_1^{(1)}c_1^{(2)}c_1^{(3)}\ket{0}_o\bra{0}\ov{c}_1^{(1)}\ov{c}_1^{(2)}\ov{c}_1^{(3)}\ket{0}_o\;,
\end{equation}
where the 2 comes form the normalization and the sign from the reordering of the fermionic operators. The upper indices label the Hilbert spaces. Equivalently in the $\mathcal{N}=1$ case

\begin{align*}
\label{final}
 \bra{0}c_1^{(1)}c_1^{(2)}&c_1^{(3)}\ov{c}_1^{(1)}\ov{c}_1^{(2)}\ov{c}_1^{(3)}c_{1/2}^{\prime\,(i)}c_{1/2}^{\prime\,(j)}\ov{c}^{\prime\,(l)}_{1/2}\ov{c}^{\prime\,(m)}_{1/2}e^{a_1\Phi^{(1)}+a_2\Phi^{(2)}+a_3\Phi^{(3)}+b_1\ov{\Phi}^{(1)}+b_2\ov{\Phi}^{(2)}+b_3\ov{\Phi}^{(3)}}\ket{0}=&\\&2\bra{0}c_1^{(1)}c_1^{(2)}c_1^{(3)}\ket{0}_o\bra{0}\ov{c}_1^{(1)}\ov{c}_1^{(2)}\ov{c}_1^{(3)}\ket{0}_o\bra{0}c_{1/2}^{\prime\,(i)}c_{1/2}^{\prime\,(j)}\ket{0}_o\bra{0}\ov{c}^{\prime\,(l)}_{1/2}\ov{c}^{\prime\,(m)}_{1/2}\ket{0}_o\\&\cdot\bra{0}e^{a_1\Phi^{(1)}+a_2\Phi^{(2)}+a_3\Phi^{(3)}+b_1\ov{\Phi}^{(1)}+b_2\ov{\Phi}^{(2)}+b_3\ov{\Phi}^{(3)}}\ket{0}_o\;.
\end{align*}
Note that in this case the signs form the bc system and the $b'c^{\prime}$ system cancel each other. The factor of 2 remains due to the chosen normalization. The indices $i,j,l,m$ take the values $1..3$, $i\neq j$, $l\neq m$ with  $\sum_{i=1}^3a_i=-2$ and $\sum_{i=1}^3b_i=-2$. Using the conservation laws described in section \ref{law}, it is possible to eliminate all creation operators except $c_1$ and $c^{\prime}_{1/2}$. Together with the current conservation, these are the only amplitudes which need to be evaluated. It is important to note that this splitting into factorized form is only possible after the conservation laws have been applied, as otherwise the signs can come out wrong.

The affine symmetry fixes the n-point function for free fields completely to be \cite{Ribault:2016sla}:
\begin{equation}
\label{amplitude}
\Braket{\prod\limits_{i=1}^nV_i}=\delta(\sum\limits_{i=1}^n\alpha_i-Q)\prod\limits_{i<j}(z_i-z_j)^{\alpha_i\alpha_j}\;,
\end{equation}
where $\alpha_i$ are the charges of the fields under the symmetry and $Q$ is the current non-conservation. The values of $Q$ and the charges of the fields are listed in table \ref{t2}. Note that for delta functions of fields, or equivalently exponential functions, the conformal weight is minus the weight of the field, such that the exponent in \eqref{amplitude} is $-\alpha_i\alpha_j$. \cite{Witten:2012bh} This allows the evaluation of all possible off-shell $n$-point functions. Note that while we have written the factorization only for the $3$-point case, it holds analogous for any $n$-point function, only the indices take values up to n instead of $3$.
Explicitly, the relevant open string amplitudes are
\begin{equation}
\bra{0}c_1(z_1)c_1(z_2)c_1(z_3)\ket{0}=(z_1-z_2)(z_1-z_3)(z_2-z_3)\;,
\end{equation}
\begin{equation}
\bra{0}c'(z_1)c'(z_2)\ket{0}=(z_1-z_2)
\end{equation}
and
\begin{equation}
\bra{0}e^{a_1\Phi^{(1)}+a_2\Phi^{(2)}+a_3\Phi^{(3)}}\ket{0}=\delta(a_1+a_2+a_3+2)\prod\limits_{i<j}(z_i-z_j)^{-a_i\cdot a_j}\;,
\end{equation}
\subsection{Picture Changing}
The evaluation of the threepoint amplitude requires the insertion of one picture changing operator(PCO) X. We choose to insert it at the location of the third insertion. Analogously for the 4-point function we insert the two required PCOs at the third and fourth puncture. The PCO has the expression\cite{Berkovits:1993xq}
\begin{equation}
X=\{Q,\xi\}=e^{\phi}(b'+c^{\prime}(T_m+\partial(c^{\prime})b')+{5\over 2}\partial^2c^{\prime})+2\partial(\eta)e^{2\phi}b+\eta\partial(e^{2\phi}b)+c\partial\xi\;.
\end{equation}
and equivalently for $\ov{X}$ by replacing all fields with there anti-holomorphic counterparts. The string field was chosen to be in the $-1$ picture in the NS sector and $-1/2$ in the R sector. For a non-vanishing amplitude in the NS sector one therefore needs to insert one $X$ and one $\ov{X}$ operator. To keep the number of punctures minimal, we insert  not the PCOs but the pictured changed versions of the string field. We define
\begin{equation}
\ov{X}XV(y)=\oint {dz\over 2\pi iz}\oint {dw\over 2\pi iw}\ov{X}(z)X(w)V(y) \;.
\end{equation}
As an example consider the tachyon field $t=c_1 \bar{c}_1 e^{-\bar{\phi }-\phi } c^{\prime}_{\frac{1}{2}} \overline{c^{\prime}}_{\frac{1}{2}}$. Its picture changed version is
\begin{align*}
\ov{X}Xt=&-\eta _{-1} e^{\phi } \bar{c}_1 c^{\prime}_{\frac{1}{2}}-\eta _{-1} e^{\phi } \bar{c}_1 c^{\prime}_{\frac{1}{2}} \overline{c^{\prime}}_{-\frac{1}{2}} \overline{c^{\prime}}_{\frac{1}{2}}-\\&c_1 \bar{\eta }_{-1} e^{\bar{\phi }} \overline{c^{\prime}}_{\frac{1}{2}}+\eta _{-1} \bar{\eta }_{-1} e^{\bar{\phi }+\phi } c^{\prime}_{\frac{1}{2}} \overline{c^{\prime}}_{\frac{1}{2}}-c_1 \bar{\eta }_{-1} e^{\bar{\phi }} c^{\prime}_{-\frac{1}{2}} c^{\prime}_{\frac{1}{2}}
   \overline{c^{\prime}}_{\frac{1}{2}}+\\&c_1 \bar{c}_1 \overline{c^{\prime}}_{-\frac{1}{2}} \overline{c^{\prime}}_{\frac{1}{2}}+c_1 \bar{c}_1 c^{\prime}_{-\frac{1}{2}} c^{\prime}_{\frac{1}{2}}+c_1 \bar{c}_1 c^{\prime}_{-\frac{1}{2}} c^{\prime}_{\frac{1}{2}} \overline{c^{\prime}}_{-\frac{1}{2}} \overline{c^{\prime}}_{\frac{1}{2}}+c_1 \bar{c}_1\;.
\end{align*}
From this expression as described earlier only the last term contributes to the bosonic amplitudes. Note that it is exactly the representation of the tachyon in the bosonic theory. But in the $\mathcal{N}=1$ theory the other terms do contribute. As this expression has indefinite ghost numbers, this allows for many different amplitudes to be non-vanishing. The tachyon is actually the only field where all 8 terms of the PCO generate a non-vanishing result.\\
Another interesting case is the open string state $c_1 \eta _{-1} c^{\prime}_{\frac{1}{2}}$. For this state holds
\begin{equation}
\oint {dz\over 2\pi i z}X(z)\;c \eta  c^{\prime}(w)=0\;.
\end{equation}
This implies that all closed string states build out of this state or its anti-holomorphic counter part have no picture changed counterpart. Therefore these fields can only appear at most quadratically in the action, even in the full non-polynomial theory. This phenomenon is quite common, out of the 138 fields 79 have no picture changed counterpart.
\subsection{B-Insertions}
For the 4-point amplitude there are not only PCOs but also $b$ insertions. These take the form \cite{Belopolsky:1994sk}
\begin{equation}
B=\sum\limits_{i=1}^4\sum\limits_{m=-1}^{\infty} B^{i}_mb^{(i)}_m+\ov{C}^{i}_m\ov{b}_m^i\;,
\end{equation}
\begin{equation}
B^{\star}=\sum\limits_{i=1}^4\sum\limits_{m=-1}^{\infty} \ov{B}^{i}_m\ov{b}^{(i)}_m+C^{i}_mb_m^i\;,
\end{equation}
with the coefficients
\begin{equation}
B^i_m=\oint {dw\over 2\pi i} {1\over w^{2+m}}{1\over f_i^{\prime}(w)}{\partial f_i(w)\over \partial \xi}\;,
\end{equation}
\begin{equation}
C^i_m=\oint {dw\over 2\pi i} {1\over w^{2+m}}{1\over f_i^{\prime}(w)}{\partial f_i(w)\over \partial \ov{\xi}}\;.
\end{equation}
The conformal maps $f_i$ are defined in section \ref{maps}. $\xi$ is the position of the unfixed puncture.
The summation runs only over the annihilation operators. Therefore they annihilate all states except the ones with the respective c-ghosts. As the states we are interested in have only $c_1$ and $c_{-1}$ oscillators, only the $m=-1$ and $m=1$ cases are needed. Additionally, only the third puncture is moved around and the insertion $z_3=\xi$ is holomorphic, such that 
\begin{equation}
C_1^i=0\;,
\end{equation}
\begin{equation}
\label{binsert}
B_1^i={1\over \rho_3}\delta_{i,3}\;.
\end{equation}
\subsection{Conservation Laws}

\label{law}
The main ingredient in calculating the interaction terms are conservation laws. We use the methods of \cite{Rastelli:2000iu}. For any primary operator $\mathcal{O}$ of conformal dimension $h$, the combination $\phi\mathcal{O}$ transforms as a holomorphic 1-form if $\phi$ has dimension $1-h$.
The correlator
\begin{equation}
\braket{\oint_{\mathcal{C}}\phi(z)\mathcal{O}(z)dzf_1\circ A_1\;f_2\circ A_2\;f_3\circ A_3}
\end{equation}
vanishes for arbitrary vertex operators $A_i$ by shrinking the contour to zero size around infinity. Therefore
\begin{equation}
\bra{V_3}\oint_{\mathcal{C}}\phi(z)\mathcal{O}(z)dz=0\;.
\end{equation}
Deforming the contour to three contours around the punctures and introducing local coordinates, this becomes
\begin{equation}
\label{conserv}
\bra{V_3}\sum_{i=1}^3\oint_{\mathcal{C}_i}dz_i\phi^{(i)}(z_i)\mathcal{O}(z_i)=0\;.
\end{equation}
The contour integration picks out the $z_i^{-1}$ terms. The operator has the mode expansion
\begin{equation}
\mathcal{O}(z_i)=\sum_{n\in \mathbb{Z}}z_i^{-n-h}\mathcal{O}_{n}\;.
\end{equation}
To eliminate the k-th mode at the $i$-th puncture a field $\phi$ with an order 1$-h+k$ pole at the puncture $i$ and less singular at the other punctures is needed. As the three punctures are mapped to 0 and $\pm\sqrt3$ the simple fields
\begin{equation}
\label{den1}
\phi_i(z)=(z-z_i)^{k+h-1}
\end{equation}
suffice for operators with conformal dimension $h>0$. Note that this expression is in the uniformizer coordinate $z$ and needs to be transformed into local coordinates before being inserted into \eqref{conserv}. For operators with $h<0$ there is a small caveat due to holomorphicity of $\phi$ at infinity. There another coordinate patch with $w=-{1\over z}$ has to be used. This imposes the condition 
\begin{equation}
\lim_{z\rightarrow\infty}z^{2h+2}\phi(z)<\infty\;.
\end{equation}
To ensure that $\phi$  is finite at infinity  the fields
\begin{equation}
\label{den2}
\phi_i=(z-z_i)^{k+h-1}\cdot\prod_{j\neq i}(z-z_j)^{-1}
\end{equation}
are used instead. Evaluating \eqref{conserv} gives conservation laws for all creation operators in terms of annihilators. For non-primary fields there are additional contributions from the anomaly in the conformal transformation. The only non-primary field used in this paper is $\partial\phi$. It transforms as
\begin{equation}
f\circ \partial\phi(z)=f'\partial\phi(f(z))-{f''(z)\over f'(z)}\;.
\end{equation}
\eqref{conserv} then becomes
\begin{equation}
\bra{V_3}\sum_{i=1}^3\oint_{\mathcal{C}_i}dz_i\phi^{(i)}(z_i)\left(\partial\phi(z_i)+{f''(z_i)\over f'(z_i)}\right)=0
\end{equation}
Note that $\phi^{(i)}$ is the scalar density used to calculate the conservation laws and $\phi$ is the field of the theory. There is no contribution from shrinking the contour around infinity due to the chosen densities. The anomaly causes the appearance of $\phi_0$ modes in the amplitudes, which effectively count the picture of the inserted state and pose no problem in the evaluation of the amplitudes.

The conservation laws obtained from the densities \eqref{den1} and \eqref{den2} replace creation operators by a sum over annihilation operators at all punctures as well as lower creation operators at the same puncture. The lowest creation operator, i.e. $\mathcal{O}_{-h}$, gets replaced only by annihilation operators, such that by iterative application of the conservation laws the amplitudes can be reduced to the basic overlap.
\subsection{Conformal Maps}
\label{maps}
The conservation laws can be calculated in terms of the coefficients of the Laurent series of the conformal maps from the worldsheets onto the sphere. For the 3 punctured sphere these are just numbers, in the 4-punctured case they are one-parameter functions. 
We follow \cite{Rastelli:2000iu} and take the convention that the 3 punctures are at $\sqrt{3}$, 0 and $\sqrt{-3}$, which avoids special treatment of a point at infinity. The punctures are labeled $z_i$ and the uniformizer $z$. The conformal maps which map the worldsheets onto the sphere are given by
\begin{align}
&z=f_n(z_n)=h\circ F_n(z_n)\\
&h(x)=-i{x-1\over x+1}\\
&F_n(x)=e^{{2\pi i\over 3}\cdot (n-1)}\cdot\left({1+ix\over 1-ix}\right)^{2/3}
\end{align}
such that $f_1(0)=\sqrt{3}$, $f_2(0)=0$ and $f_3(0)=-\sqrt{3}$. The series expansions around the punctures are
\begin{align*}
f_1 (z_1) =& \sqrt {3}+{\frac {8}{3}}\,z_1
+{\frac {16}{9}}\,\sqrt {3}\,{z_1}^{2}+{\frac
{248}{81}}\,{z_1}^{3}+{\frac {416}{243}}\,
\sqrt {3}\,{z_1}^{4}+{\frac {2168}{
729}}\,{z_1}^{5}+O\left ({z_1}^{6}\right )\,.\\
f_2 (z_2)=& 0+ 
{\frac {2}{3}}\, z_2-{\frac {10}{81}}\, {z_2}^{3}
+{\frac {38}{729}}\,{z_2}^{5}+O\left
({z_2}^{7}\right )\,.
 \\
f_3 (z_3) =&
-\sqrt {3}+{\frac {8}{3}}\, z_3-{\frac {16}{9}}\,\sqrt
{3}\, {z_3}^{2}+{\frac {248}{81}}\, {z_3}^{3}-{\frac {416}{243}}\,\sqrt
{3}\, {z_3}^{4} +{\frac {2168}{
729}}\, {z_3}^{5}+O({z_3}^{6})  \,.
\end{align*}

For the 4-punctured sphere we adopt the convention that the punctures are at $0$, $1$, $\xi$ and $\infty$ to be able to use the fits provided by Moeller in \cite{Moeller:2004yy}. The conformal maps are calculated using Strebel differentials. For a 4-punctured sphere the general form is
\begin{equation}
\phi(z) = {-(z^2 - \xi)^2 \over z^2 \, (z-1)^2 \, (z-\xi)^2} + 
{a \over z \, (z-1) \, (z-\xi)} \;.
\end{equation}
Here $a$ is a complex number which depends on the position of the puncture $\xi$. The strebel condition, i.e. the condition that all critical trajectories close, suffices to fix $a$ uniquely. Expanding this expression around the punctures and comparing it to the known expression in local coordinates
\begin{equation}
\phi(w_i)={-1\over w_i^2}
\end{equation}
allows to determine the coefficients of the conformal transformations.
We expand them as 
\begin{equation}
\label{expansion}
f_i(z_i)=z_i+\rho_i z_i+\sum\limits_{n=2}^{\infty} d_{i,n-1}(\rho_iz_i)^n\;.
\end{equation}

\begin{figure}[h]
  \centering
  \includegraphics[scale=0.5]{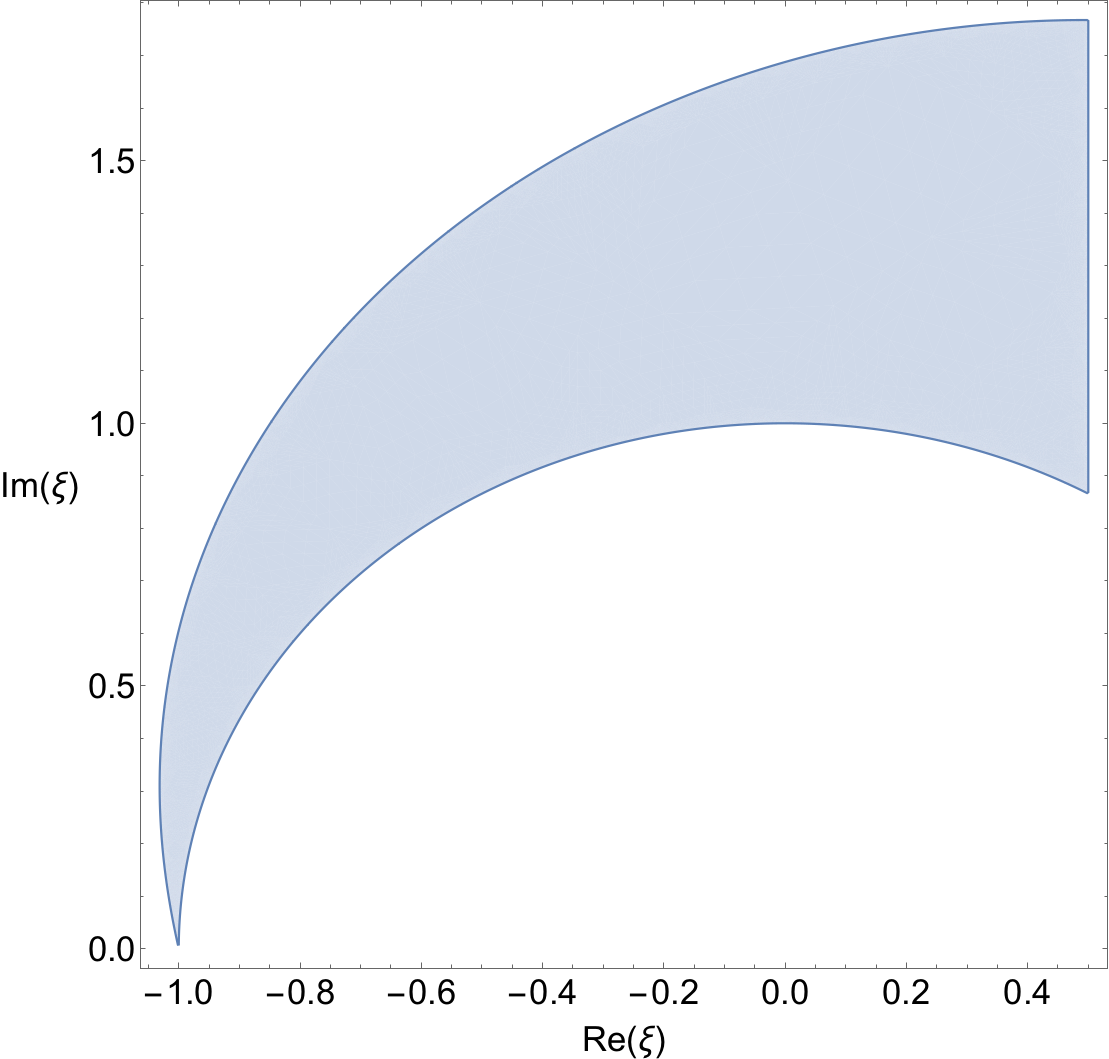}
  \caption{Fundamental domain of integration $\mathcal{A}$}
  \label{region}
\end{figure}

\cite{Moeller:2004yy} gives fits for $a$ as well as for the mapping radii $\rho_i$. Finally, one needs the area of integration. One can use symmetries of the system to map into a fundamental region $\mathcal{A}$ defined by the conditions 
\begin{equation}
\text{Re}[\xi]\le 1/2\qquad \text{Im}[\xi]\ge 0\qquad 1\le\text{Abs}[\xi]\le r(\theta)
\end{equation}
where $r(\theta)$ is a function given as a fit in \cite{Moeller:2004yy}. Figure \ref{region} shows the fundamental domain. The rest of the moduli space is obtained by the PSL$(2,\mathbb{C})$transformations $\xi\rightarrow 1-\xi$, $\xi\rightarrow 1/\xi$ as well as complex conjugation $\xi\rightarrow\ov{\xi}$ and combinations thereof.
\subsection{Sample Calculations of Quartic Terms}
\label{terms}
Here we will evaluate exemplary two of the quartic terms. The first is the four tachyon amplitude. Due to b'c' conservation, from the picture changed tachyon only the $c_1\ov{c}_1$ term contributes. As there are only $c_1$ modes in the tachyon, only the $b_{-1}$ insertion contributes, which due to \eqref{binsert} appears only at the third puncture and annihilates the picture changed tachyon. Therefor,
\begin{equation*}
\braket{t,t,t,t}={2\over \pi}\int_{V_{0,4}} dxdy {1\over \rho_3^2}\bra{\Sigma}c_1\ov{c}_1c^{\prime}_{1/2}\ov{c}^{\prime}_{1/2}e^{-\Phi-\ov{\Phi}},c_1\ov{c}_1c^{\prime}_{1/2}\ov{c}^{\prime}_{1/2}e^{-\Phi-\ov{\Phi}},1,c_1\ov{c}_1\ket{0}
\end{equation*}
The fields at puncture 1,2, and 4 are all primaries of weight -2, giving a factor of ${1/ \rho^2}$ each, such that
\begin{equation}
\braket{t,t,t,t}={2\over \pi}\int_{V_{0,4}}dxdy{1\over \rho_1^2\rho_2^2\rho_3^2\rho_4^2}={24\over \pi}\int_{\mathcal{A}}dxdy{1\over \rho_1^2\rho_2^2\rho_3^2\rho_4^2}\approx 72.55\;.\footnote{The value given here is the result of our computation. There is a slight missmatch between this value of 72.55 and the value given in e.g \cite{Moeller:2004yy} of 72.41. We used the NIntgerate function of Mathematica 11.3, while they used a Monte Carlo technique. As this value has to be divided by 24 to obtain the coefficient in the action this difference is marginal and has no effect on the solutions.}
\end{equation}
Here we can simply take 12 times the value of the fundamental domain as the integrand is invariant under the exchange of punctures.
As a second example we take the $t^2B_{11}B_{16}$ term. $XB_{11}$ vanishes and $X\ov{X}B_{16}=25c'\ov{c}'$ has the wrong ghost number, such that the only contributing terms are $\braket{B_{11},B_{16},t,t}$ and $\braket{B_{16},B_{11},t,t}$. I.e. we have to evaluate
\begin{equation}
\braket{B_{11},B_{16},t,t}=\braket{e^{-2 \phi }e^{-2 \bar{\phi }}c_1\text{c}^{\prime}_{\frac{1}{2}}\xi _{-1}\bar{c}_1\overline{\text{c}}^{\prime}_{\frac{1}{2}}\bar{\xi }_{-1},c_1\text{c}^{\prime}_{\frac{1}{2}}\eta _{-1}\bar{c}_1\overline{\text{c}}^{\prime}_{\frac{1}{2}}\bar{\eta }_{-1},1,c_1\bar{c}_1}
\end{equation}

Using the conservation laws for $\xi_{-1}$, the only term contributing is the insertion of a $\xi_1$ in the second puncture with coefficient $-\rho _1 \rho _2 d_{1,1} d_{2,1}$. The same holds for the $\ov{\xi}_{-1}$ mode. Therefore
\begin{equation}
\braket{B_{11},B_{16},t,t}={2\over \pi}\int_{V_{0,4}}dxdy{1\over \rho_3^2}\rho^2 _1 \rho^2 _2 d^2_{1,1} d^2_{2,1}\braket{e^{-2 \phi }e^{-2 \bar{\phi }}c_1\text{c}^{\prime}_{\frac{1}{2}}\bar{c}_1\overline{\text{c}}^{\prime}_{\frac{1}{2}},c_1\text{c}^{\prime}_{\frac{1}{2}}\bar{c}_1\overline{\text{c}}^{\prime}_{\frac{1}{2}},1,c_1\bar{c}_1}
\end{equation}

\begin{equation}
\label{step2}
={2\over \pi}\int_{V_{0,4}}dxdy{d_{1,1}^2d_{2,1}^2\over \rho_1\rho_2\rho_3^2\rho_4^2}\;.
\end{equation}
The $d_{i,j}$ are the expansion coefficients of the conformal transformations, see \eqref{expansion}.
This time the integrand is not invariant, such that we have to sum over the different contributions. For this we apply the transformations in the appendix of \cite{Yang:2005ep}. There are in total 6 regions,
\begin{equation}
\mathcal{A},\quad 1-\mathcal{A},\quad {1\over \mathcal{A}},\quad 1-{1\over \mathcal{A}},\quad {1\over 1-\mathcal{A}}\quad \text{and}\quad {1\over 1-{1\over\mathcal{A}}}\;.
\end{equation}
 These 6 regions can be mapped back to region $\mathcal{A}$ using the transformations
\begin{equation}
g_1=\xi\rightarrow{1\over \xi},\qquad g_2=\xi\rightarrow{1-\xi}
\end{equation}
The volume element and derivatives are invariant under $g_2$, while for $g_1$
\begin{equation}
dxdy\rightarrow {dxdy\over |\xi|^4},\qquad \partial_{\xi}\rightarrow \xi^2\partial_{\xi},\qquad \ov{\partial}_{\xi}\rightarrow \ov{\xi}^2\ov{\partial}_{\xi}
\end{equation}
Using this, \eqref{step2} becomes
\begin{align*}
&{4\over \pi}\int\limits_{\mathcal{A}}dxdy\Bigl(\frac{2 d_{1,1}^2 d_{2,1}^2}{\rho _1 \rho _2 \rho _3^2 \rho _4^2}+\frac{\left(d_{1,1}+1\right){}^2 \left(d_{4,1}+1\right){}^2}{\rho _1 \rho _2^2 \rho _3^2 \rho _4}+\frac{2 \left(1-d_{2,1}\right){}^2 d_{4,1}^2}{\rho _1^2 \rho _2 \rho _3^2 \rho _4}\\&+\frac{\left(d_{1,1}+1\right){}^2 \left(d_{4,1}+1\right){}^2}{\rho _1 \rho _2^2 \rho _3^2 \rho _4}\Bigr)\approx 4637.76\;.
\end{align*}
 the overall factor of two comes from the complex conjugation symmetry. The Evaluation of $\braket{B_{16},B_{11},t,t}$ proceeds along similar lines with the roles of $\xi$ and $\eta$ exchanged and with the same result.

\subsection{Complete Action}
\label{data}
In the source files there is a Mathematica file "Potential.m". It includes a single list with two entries. The first entry is the potential itself in terms of 138 fields $A_i$. The values for the cubic terms are given as ratios of integers. The quartic coefficients are given to 10 digits precision, but note that the numerical errors allow only for 2 digits precision.  The second entry is a dictionary identifying the fields $A_i$ in the action with the states of the CFT. For example $A_{27}$ corresponds to the tachyon. The numeration of the fields is arbitrary and follows no special ordering. 
\clearpage
\bibliography{references}  

\providecommand{\href}[2]{#2}\begingroup\raggedright\begin{thebibliography}{10}

\bibitem{Adams:2001sv}
A.~Adams, J.~Polchinski, and E.~Silverstein, ``{Don't panic! Closed string
  tachyons in ALE space-times},'' {\em JHEP} {\bf 10} (2001) 029,
\href{http://www.arXiv.org/abs/hep-th/0108075}{{\tt hep-th/0108075}}.

\bibitem{David:2001vm}
J.~R. David, M.~Gutperle, M.~Headrick, and S.~Minwalla, ``{Closed string
  tachyon condensation on twisted circles},'' {\em JHEP} {\bf 02} (2002) 041,
\href{http://www.arXiv.org/abs/hep-th/0111212}{{\tt hep-th/0111212}}.

\bibitem{Headrick:2004hz}
M.~Headrick, S.~Minwalla, and T.~Takayanagi, ``{Closed string tachyon
  condensation: An Overview},'' {\em Class. Quant. Grav.} {\bf 21} (2004)
  S1539--S1565,
\href{http://www.arXiv.org/abs/hep-th/0405064}{{\tt hep-th/0405064}}.

\bibitem{Suyama:2005wd}
T.~Suyama, ``{Closed string tachyon condensation in supercritical strings and
  RG flows},'' {\em JHEP} {\bf 03} (2006) 095,
\href{http://www.arXiv.org/abs/hep-th/0510174}{{\tt hep-th/0510174}}.

\bibitem{Hellerman:2004zm}
S.~Hellerman, ``{On the landscape of superstring theory in D $>$ 10},''
\href{http://www.arXiv.org/abs/hep-th/0405041}{{\tt hep-th/0405041}}.

\bibitem{Hellerman:2004qa}
S.~Hellerman and X.~Liu, ``{Dynamical dimension change in supercritical string
  theory},''
\href{http://www.arXiv.org/abs/hep-th/0409071}{{\tt hep-th/0409071}}.

\bibitem{Hellerman:2006ff}
S.~Hellerman and I.~Swanson, ``{Dimension-changing exact solutions of string
  theory},'' {\em JHEP} {\bf 09} (2007) 096,
\href{http://www.arXiv.org/abs/hep-th/0612051}{{\tt hep-th/0612051}}.

\bibitem{Hellerman:2006hf}
S.~Hellerman and I.~Swanson, ``{Cosmological unification of string theories},''
  {\em JHEP} {\bf 07} (2008) 022,
\href{http://www.arXiv.org/abs/hep-th/0612116}{{\tt hep-th/0612116}}.

\bibitem{West:2002hh}
P.~C. West, ``{The Spontaneous compactification of the closed bosonic
  string},'' {\em Phys. Lett.} {\bf B548} (2002) 92--96,
\href{http://www.arXiv.org/abs/hep-th/0208214}{{\tt hep-th/0208214}}.

\bibitem{Lust:1987ik}
D.~Lüst, ``{Covariant Fermionic and Heterotic Strings From the Bosonic
  String},'' {\em Nucl. Phys.} {\bf B292} (1987)
381--399.

\bibitem{PhysRevD.42.1289}
V.~A. Kosteleck\'y and S.~Samuel, ``Collective physics in the closed bosonic
  string,'' {\em Phys. Rev. D} {\bf 42} (Aug, 1990) 1289--1292.

\bibitem{Moeller:2004yy}
N.~Moeller, ``{Closed bosonic string field theory at quartic order},'' {\em
  JHEP} {\bf 11} (2004) 018,
\href{http://www.arXiv.org/abs/hep-th/0408067}{{\tt hep-th/0408067}}.

\bibitem{Belopolsky:1994bj}
A.~Belopolsky, ``{Effective Tachyonic potential in closed string field
  theory},'' {\em Nucl. Phys.} {\bf B448} (1995) 245--276,
\href{http://www.arXiv.org/abs/hep-th/9412106}{{\tt hep-th/9412106}}.

\bibitem{Yang:2005rx}
H.~Yang and B.~Zwiebach, ``{A Closed string tachyon vacuum?},'' {\em JHEP} {\bf
  09} (2005) 054,
\href{http://www.arXiv.org/abs/hep-th/0506077}{{\tt hep-th/0506077}}.

\bibitem{Moeller:2006cv}
N.~Moeller and H.~Yang, ``{The Nonperturbative closed string tachyon vacuum to
  high level},'' {\em JHEP} {\bf 04} (2007) 009,
\href{http://www.arXiv.org/abs/hep-th/0609208}{{\tt hep-th/0609208}}.

\bibitem{Moeller:2006cw}
N.~Moeller, ``{Closed Bosonic String Field Theory at Quintic Order:
  Five-Tachyon Contact Term and Dilaton Theorem},'' {\em JHEP} {\bf 03} (2007)
  043,
\href{http://www.arXiv.org/abs/hep-th/0609209}{{\tt hep-th/0609209}}.

\bibitem{Berkovits:1993xq}
N.~Berkovits and C.~Vafa, ``{On the Uniqueness of string theory},'' {\em Mod.
  Phys. Lett.} {\bf A9} (1994) 653--664,
\href{http://www.arXiv.org/abs/hep-th/9310170}{{\tt hep-th/9310170}}.

\bibitem{Blumenhagen:2013fgp}
R.~Blumenhagen, D.~Lüst, and S.~Theisen, {\em {Basic concepts of string
  theory}}.
\newblock Theoretical and Mathematical Physics. Springer, Heidelberg, Germany,
2013.
\newblock

\bibitem{deLacroix:2017lif}
C.~de~Lacroix, H.~Erbin, S.~P. Kashyap, A.~Sen, and M.~Verma, ``{Closed
  Superstring Field Theory and its Applications},'' {\em Int. J. Mod. Phys.}
  {\bf A32} (2017), no.~28n29, 1730021,
\href{http://www.arXiv.org/abs/1703.06410}{{\tt 1703.06410}}.

\bibitem{Yang:2005ep}
H.~Yang and B.~Zwiebach, ``{Dilaton deformations in closed string field
  theory},'' {\em JHEP} {\bf 05} (2005) 032,
\href{http://www.arXiv.org/abs/hep-th/0502161}{{\tt hep-th/0502161}}.

\bibitem{Kudrna:2018mxa}
M.~Kudrna and M.~Schnabl, ``{Universal Solutions in Open String Field
  Theory},''
\href{http://www.arXiv.org/abs/1812.03221}{{\tt 1812.03221}}.

\bibitem{Headrick:2018dlw}
M.~Headrick and B.~Zwiebach, ``{Minimal-area metrics on the Swiss cross and
  punctured torus},''
\href{http://www.arXiv.org/abs/1806.00450}{{\tt 1806.00450}}.

\bibitem{Headrick:2018ncs}
M.~Headrick and B.~Zwiebach, ``{Convex programs for minimal-area problems},''
\href{http://www.arXiv.org/abs/1806.00449}{{\tt 1806.00449}}.

\bibitem{Naseer:2019zau}
U.~Naseer and B.~Zwiebach, ``{Extremal isosystolic metrics with multiple bands
  of crossing geodesics},''
\href{http://www.arXiv.org/abs/1903.11755}{{\tt 1903.11755}}.

\bibitem{Moosavian:2017fta}
S.~F. Moosavian and R.~Pius, ``{Hyperbolic Geometry of Superstring Perturbation
  Theory},''
\href{http://www.arXiv.org/abs/1703.10563}{{\tt 1703.10563}}.

\bibitem{Moosavian:2017sev}
S.~F. Moosavian and R.~Pius, ``{Hyperbolic Geometry and Closed Bosonic String
  Field Theory II: The Rules for Evaluating the Quantum BV Master Action},''
\href{http://www.arXiv.org/abs/1708.04977}{{\tt 1708.04977}}.

\bibitem{Moosavian:2017qsp}
S.~F. Moosavian and R.~Pius, ``{Hyperbolic Geometry and Closed Bosonic String
  Field Theory I: The String Vertices Via Hyperbolic Riemann Surfaces},''
\href{http://www.arXiv.org/abs/1706.07366}{{\tt 1706.07366}}.

\bibitem{Hellerman:2007ym}
S.~Hellerman and I.~Swanson, ``{Supercritical N = 2 string theory},''
\href{http://www.arXiv.org/abs/0709.2166}{{\tt 0709.2166}}.

\bibitem{Yang:2005iua}
H.-t. Yang and B.~Zwiebach, ``{Testing closed string field theory with marginal
  fields},'' {\em JHEP} {\bf 06} (2005) 038,
\href{http://www.arXiv.org/abs/hep-th/0501142}{{\tt hep-th/0501142}}.

\bibitem{Ribault:2016sla}
S.~Ribault, ``{Minimal lectures on two-dimensional conformal field theory},''
  in {\em {SciPost Phys. Lect. Notes 1 (2018)}}.
\newblock 2016.
\newblock
\href{http://www.arXiv.org/abs/1609.09523}{{\tt 1609.09523}}.
\newblock

\bibitem{Witten:2012bh}
E.~Witten, ``{Superstring Perturbation Theory Revisited},''
\href{http://www.arXiv.org/abs/1209.5461}{{\tt 1209.5461}}.

\bibitem{Belopolsky:1994sk}
A.~Belopolsky and B.~Zwiebach, ``{Off-shell closed string amplitudes: Towards a
  computation of the tachyon potential},'' {\em Nucl. Phys.} {\bf B442} (1995)
  494--532,
\href{http://www.arXiv.org/abs/hep-th/9409015}{{\tt hep-th/9409015}}.

\bibitem{Rastelli:2000iu}
L.~Rastelli and B.~Zwiebach, ``{Tachyon potentials, star products and
  universality},'' {\em JHEP} {\bf 09} (2001) 038,
\href{http://www.arXiv.org/abs/hep-th/0006240}{{\tt hep-th/0006240}}.

\end{thebibliography}\endgroup
\bibliographystyle{utphys}
\end{document}